\documentclass[11pt]{article}
\usepackage{hyperref}
\pdfoutput=1

\begin{document}
\title{Excited Sessile Drops Perform Harmonically}
\author{Chun-Ti Chang$^1$, Susan Daniel$^{1,2}$, Paul H. Steen$^{1,2}$\\ \vspace{6pt}\\
$^1$Theoretical and Applied Mechanics,\\
$^2$Chemical and Biomolecular Engineering,
\\ Cornell University, Ithaca, NY 14853, USA}
\maketitle

\begin{abstract}
In our fluid dynamics video, we demonstrate our method of visualizing and identifying various mode shapes of mechanically oscillated sessile drops. By placing metal mesh under an oscillating drop and projecting light from below, the drop's shape is visualized by the visually deformed mesh pattern seen in the top view. The observed modes are subsequently identified by their number of layers and sectors. An alternative identification associates them with spherical harmonics, as demonstrated in the tutorial. Clips of various observed modes are presented, followed by a 10-second quiz of mode identification.
\end{abstract}
% main text
\section{Introduction}
Two sample videos are \href{TBD}{DanceDropHiResol.mp4} and \href{TBD}{DanceDropLoResol.mp4}.\\ \\
%-
In our project, we report the dynamics of mechanically-excited sessile drops by characterizing their resonant frequencies and corresponding mode shapes. With our fluid dynamics video, we present the most qualitative portion of the project: technique of visualization and scheme of mode identification. For frequency response and comparison to theory, please refer to \emph{Substrate constraint modifies the Rayleigh spectrum of vibrating sessile drops}, Physical Review E 88, 023015 (2013).\\ \\
%-
In the experiment, a liquid drop is deposited on a clear glass substrate and oscillated in a plane-normal direction with a mechanical vibrator. To visualize shapes of the oscillating drop, a metal mesh is placed under the glass substrate. Light is projected upward from below the mesh. As the light passes through the mesh and the drop, the drop's surface refracts the light according to its deformation. The refraction produces visual deformation of the mesh from the top view, thereby visualizing the shape deformation of the drop in a human-readable way.\\ \\
%-
Given a particular mode, we identify its shape with a visual approach: we count from the top view the number of sectors, which is equal to $l$, and from the side view the number of layers $n$. Alternatively, an analogy is drawn between these shapes and spherical harmonics:
\[
  r_{kl} (\theta, \psi) = 1 + \epsilon P_k^l (\cos \theta) \cos (l \psi)
\]
where $r_{kl}$ is the radial coordinate of any point Q on the surface of a drop oscillating at its $(k, l)$ mode, $\theta$ and $\psi$ the polar and azimuthal angles of Q, $\epsilon \ll 1$ the size of disturbance, $P_k^l$ the Legendre function of degree $k$ and order $l$. The expression describes a unit hemisphere deformed by the family of spherical harmonic disturbance. These modes are classified as zonal ($l = 0$), sectoral ($k = l \neq 0$) and tesseral ($k > l > 0$). Although specifying $k$ and $l$ fully prescribes a mode mathematically, the visual approach is proposed with the hope of facilitating mode identification in the most physically intuitive way possible. The number of layers $n$ relates to $k$ and $l$ as
\[
  n = \frac{k - l}{2} + 1
\]
Among the three types of modes, zonal modes are those only with layers, sectorals only with sectors in one layer, and tesserals with both layers and setors. A full catalog of all observed modes and a gallery of selected clips for (non-axisymmetric) modes follow the mode-identification tutorial and leads to a 10-second quiz, with the hope of engaging the audience into the video content.
%-
%\begin{thebibliography}{1}
%
%\bibitem{ourPRE}
% Chun-Ti Chang, Joshua B. Bostwick, Paul H. Steen, Susan Daniel,
% \emph{Substrate constraint modifies the Rayleigh spectrum of vibrating sessile drops},
% Physical Review E 88, 023015 (2013)
%
%\end{thebibliography}
%-
\end{document}